\documentclass[3p,times]{elsarticle}

\usepackage{amssymb}
\usepackage{hyperref}
\usepackage{amsmath}
\usepackage{amsthm}
\usepackage{amsfonts}
\usepackage{ mathrsfs }
\usepackage{xcolor}
\usepackage[utf8]{inputenc}
\usepackage{graphicx,subfig}
\newcommand{\be}{\begin {equation}}
\newcommand{\ee}{\end{equation}}
\newcommand{\bee}{\begin {equation*}}
\newcommand{\eee}{\end{equation*}}

\newcommand{\p}{\partial}

\newtheorem{prop}{Proposition}

\journal{Physica D}

\begin{document}

\begin{frontmatter}



\title{The impact of a wind switch on the stability of traveling fronts in a reaction-diffusion model of fire propagation }


\author[inst1,inst4]{Olivia Chandrasekhar\corref{corauthor}}

\cortext[corauthor]{Corresponding author at Los Alamos National Laboratory, Los Alamos, NM, 87544, USA. \textit{Email address}: ochandrasekhar@lanl.gov.}

\affiliation[inst1]{organization={Department of Mathematics, University of North Carolina at Chapel Hill},
            city={Chapel Hill, NC}, 
            postcode={27599}, 
            country={USA}}

\author[inst2]{C.K.R.T. Jones}

\affiliation[inst2]{organization={Department of Mathematics, George Mason University},
            city={Fairfax, VA },
            postcode={22030},
            country={USA}}

\author[inst3]{Blake Barker}
\affiliation[inst3]{organization={Department of Mathematics, Brigham Young University},
            city={Provo, UT},
             postcode={84602},
            country={USA}}

\author[inst4]{Rodman Linn}

\affiliation[inst4]{organization={Earth and Environmental Sciences Division},
            addressline={Los Alamos National Laboratory}, 
            city={Los Alamos, NM},
            postcode={87544}, 
            country={USA}}

\begin{abstract} \par

For certain values of the wave speed parameter, evolution equations for the temperature of a region of fuel admit traveling wave solutions describing fire fronts. We consider such a system in the form of a nonlinear reaction-diffusion equation with a first-order forcing term capturing the combined effects of ambient and fire-induced wind. The fire-induced wind is introduced by way of a piecewise continuous function that ``switches" in space. \par 
We demonstrate that, in the case of a spatially dependent wind, traveling wave solutions corresponding to fire fronts exist for a continuum of wave speeds rather than for a single unique speed. Using geometric methods, we determine the range of allowable speeds, refine this range to only those fronts which will persist in nature, and develop a selection mechanism to identify the specific wind configuration corresponding to the most stable solution. For this spectrally preferred front, we find that the wind switch occurs ahead of the fireline in a manner consistent with the physics of air entrainment. Even when the wind is not coupled to the temperature and is instead imposed as an external forcing, the conditions on the existence and stability of front solutions force the wind term in to a configuration reflective of physical reality.

\end{abstract}

\end{frontmatter}


\section{Introduction}\label{introduction}
Understanding the spatial structure of traveling fire fronts is key to characterizing their behavior. This is true for fires at all scales, including wildfires, prescribed fires, and laboratory-scale burns. Mathematical models of fire propagation describing only the interaction between key physical quantities---such as temperature, fuel and wind--- succinctly capture the dynamics and structure of the underlying physical system in a way that lends itself to rigorous mathematical analysis. Using such an approach allows us describe the complex nonlinear processes that contribute to observed behaviors and increases our ability to anticipate emergent phenomena.\par
In this work, we consider an example of a nonlinear reaction-diffusion equation for the evolution of the temperature of a region of homogeneous fuel. Partial differential equations of this form have been used as a minimal model for combustion in a variety of settings (including \cite{mandel2008wildland}, \cite{norbury1988travelling1}, \cite{norbury1988travelling2} and \cite{ghazaryan2015stability}) in part because they admit traveling wave solutions that may be used to model propagating fire fronts. We impose a spatially dependent advection term as a first-order forcing capturing the wind velocity in the vicinity of the front. Using a dynamical systems framework, we address both the existence and stability of these front solutions for qualitatively different wind configurations. We establish the concept of a spectrally preferred front by finding the minimum of the largest stable eigenvalue across all solution possibilities. Our results show that the preferred front only exists for a wind configuration consistent with the physics dictating the behavior of the wind field in the vicinity of a fire.  \par
  A particularly important aspect of spatial and temporal analysis of fire behavior is capturing the evolution of the fireline as it propagates, where the fireline is defined as the boundary between regions of burning and unburned fuel \cite{hilton2018incorporating}. The evolution of the fireline as a two-dimensional geometric object has been studied using, for example, level-set methods (see \cite{lautenberger2013wildland}, \cite{mallet2009modeling}). There has also been work done to understand how local processes and environmental factors, such as fuel heterogeneity \cite{atchley2021effects} and wind conditions \cite{Linn2005numerical}, influence its shape. Generally, it is the structure of the front in the transverse direction---orthogonal to the motion of the fire---that is under consideration. See, for example, \cite{canfield2014numerical} for a discussion of the two-way feedback between the shape and curvature of the fireline and its dynamics in simulated grassfires. \par
\begin{figure*}
    \centering
    \includegraphics[scale=0.9]{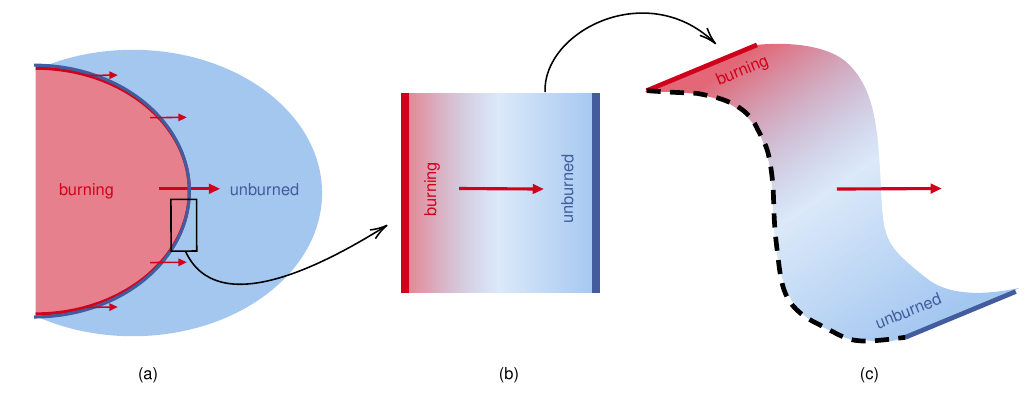}
    \caption{The fireline depicted as the boundary between regions of burning and unburned fuel in (a), then zoomed in to illustrate the transition zone between burning and unburned fuel in (b), then viewed again as a two-dimensional object in (c). The profile solutions we consider in this work correspond to the dotted black line in figure (c), asympotically connecting regions of maximum and ambient temperature (corresponding to regions of burning and unburned fuel). The direction of motion of the fireline and the traveling wave is indicated by the red arrow(s) across all three figures.}
    \label{fireline_fig}
\end{figure*}

 
The structure of the fire in the streamwise direction---parallel to the direction of motion---has been explored much less, although it is known that the thickness of the fireline in this direction influences the movement of the front. This thickness  depends on both the speed of propagation and the residence time, or time to burn,  of the fuel. In scenarios where the residence time is much larger than the ignition time---in other words, when the process of ignition occurs on a much faster time scale than the process of fuel consumption---the spatial extent of the transition zone is small. Such scenarios include most wildfires and prescribed fires (in contrast to, for example, detonations). When the timescales of the problem are separated in this way, the fireline can be conceptualized as a steep temperature profile corresponding to the initial ignition followed by a slowly declining tail as the cooling process begins to dominate.  It is the initial region of transition, consisting of a steep gradient in the temperature, that we focus on in this work. See fig. \ref{fireline_fig} for a schematic diagram illustrating this region in the context of a two-dimensional fireline.

The fact that fire generates a local wind field, often referred to as the entrainment field, as the result of entrainment of cool air ahead of the fireline is well known: see \cite{nelson2012entrainment} for an in-depth description of various wind regimes in which this occurs, \cite{linn2012using} for a discussion of the phenomenon in coupled fire-atmosphere simulations and \cite{dipierro2024simple} for a mathematical treatment of the topic. The generation of this local velocity field (referred to as ``pyrogrenic flow" by \cite{dipierro2024simple} and ``fire-induced wind" throughout much of this paper) is due to the buoyancy of the air heated by the combustion reaction and the principle of conservation of mass. Specifically, as air heated by combustion and burning fuel ahead of the fireline rises, cold air is pulled in to replace it. \par  
For a straight fireline, the strongest effects from entrainment are in front of and behind the front, with the entrainment ahead of the front making a contribution to the overall velocity field in opposition to the ambient wind. In wind-driven fires, where the ambient wind is much larger than the local, fire-induced wind, a wind term that only captures the ambient wind is a reasonable estimation of reality. As a first-order approximation, this wind field may be thought of as constant in space. However, in regimes where the environmental and fire-induced wind are on the same order of magnitude, a wind term that changes in value in space is necessary to capture the entrainment field and its interactions with the ambient wind. \par

The implication of this three-dimensional entrainment phenomenology in a one-dimensional problem is that, in the vicinity of the traveling front, the local velocity field in the streamwise direction changes in magnitude and, potentially, reverses direction. We refer to this behavior as ``switching" in the wind field. In this paper, we consider a generalized spatially dependent wind term that captures this switching behavior and explore how the constraints on existence and stability of solutions to a minimal model of fire propagation dictate the wind configuration that corresponds to the mathematically preferred solution. \par
Specifically, we consider a one-dimensional, nonlinear reaction-diffusion equation for the temperature of a region of burning fuel. The existence and stability of traveling wave solutions in the constant wind case to equations of this form is well understood: see \cite{sandstede2002stability} for the general theory. For a spatially dependent wind term, these issues are less well understood. Through a combination of geometric arguments and numerical calculations, we demonstrate the existence of traveling front solutions for a range of wave speed values dependent on the wind regime parameters. We further demonstrate that the stability properties of these solutions depend on the qualitative nature of the wind field. \par
These results give us tools to better understand the properties of the solution that will persist in nature in a system with a spatially dependent wind. We define the spectrally preferred  front as the solution for which the corresponding largest eigenvalue is minimized and identify the unique speed at which this preferred front travels as well as the spatial location of the fire front relative to the wind switch. These findings allow us to identify, for a given wind configuration, the expected distance between and relative location of the fireline and the wind switch. We find that, for the spectrally preferred solution, the wind switch occurs ahead of the fireline. This reflects the physical motivation described above: propagating fires will entrain cool air ahead of the fireline. It is noteworthy that, despite not modeling the wind or coupling it to the temperature, this minimal, temperature-based model of fire propagation is able to describe a physically relevant fire-induced wind. \par
\section{The model}
We begin with the following model, modified from \cite{mandel2008wildland}, for the evolution of the temperature $u(x,t)$ of the fire layer, consisting of a fixed, one-dimensional region of homogeneous fuel and the layer of air immediately above it.
\be\label{nondimensional equation}
\frac{\p u}{\p t}= \frac{\p^2 u}{\p x^2}-w \frac{\p u}{\p x} +v^*r(u)-lu 
\ee
This model has elements of diffusion due to radiative heat transfer, advection due to the local wind velocity, a nonlinear reaction term representing combustion, and loss of heat to the atmosphere due to convective heat transfer. We assume this latter effect dominates radiative heat transfer, resulting in the linear form of the loss term $-lu$, where $l$ dictates the rate of heat loss to the surrounding environment. The parameter $v^* \in [0,1]$ represents the fraction of the potential fuel load available at the beginning of the reaction and the advection coefficient $w$ is the wind velocity term. \par 
The piecewise continuous, Arrhenius-type source term ties the combustion reaction to the temperature of the fire layer and is defined as 
\be \label{source term}
r(u)=
\begin{cases}
e^{-1/u} \: \: &u>0 \\
0 \: \: &u \leq 0
\end{cases}
\ee

\section{Traveling wave solutions}\label{traveling wave solutions}
Systems of the form \eqref{nondimensional equation} are known to admit traveling front solutions connecting stationary states of the underlying system. In this model, these front solutions connect regions of high temperature to regions of ambient, or zero, temperature. As such, they may be used as a proxy to study the propagating fire fronts seen in all manner of experimental, natural and simulated fires. \par
To center ourselves in the frame of reference of the front solutions, we convert to traveling coordinates through the change of variable $z=x-ct$ where $c$ is the speed of the traveling wave, resulting in 
\begin{align}\label{traveling equation}
    \begin{split}
    &\frac{\p u}{\p t} =\frac{\p^2 u}{\p z^2}+(c-w(z))\frac{\p u}{\p z} +v^*r(u)-lu\\
    \end{split}    
\end{align}
Note that we have introduced spatial dependence to the wind term in the frame of the fire front. The equation governing front solutions, which have spatial structure but do not change in time, is given by 
\be\label{reduced TW equation}
0=\frac{\p^2 u}{\p z^2}+(c-w(z))\frac{\p u}{\p z}+v^*r(u)-lu
\ee
Solutions to \eqref{reduced TW equation} are traveling waves connecting regions of ambient temperature to regions of maximum temperature, where the maximum temperature is determined as a function of the initial fuel load $v^*$.
As a first order system, \eqref{reduced TW equation} becomes
\begin{align}\label{first order system}
\begin{split}
  &u'=s \\ 
 & s'=-(c-w(z))s-v^*r(u)+lu 
 \end{split}
\end{align}
Fixed points of this system occur at zeros of the reaction term 
\be\label{reaction term}
f(u,v^*)=v^*r(u)-lu
\ee which has a quasi-cubic structure reflective of the underlying bistability of the problem.  For values of $v^*$ above some threshold, $f(u,v^*)$ has three roots: $u_0$, the ambient temperature (zero for the nondimensionalized system), $u_1$, the ``auto-ignition" temperature, representing the temperature above which the reaction becomes self-sustaining, and $u_2,$ the maximum temperature the system can reach given the initial fuel load $v^*$. \par
For notational clarity, in what follows we will omit the parameter $v^*$ in the statement of \eqref{reaction term} and simply write $f(u)$ to denote the reaction term.

\begin{figure}[h]
\centering
\includegraphics[scale=.55]{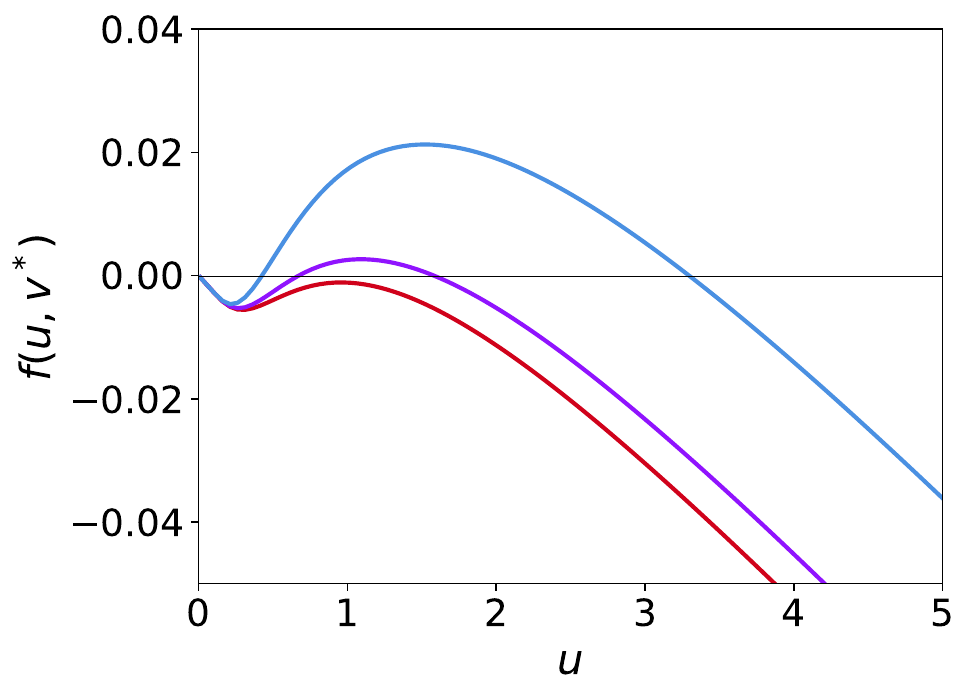}
\caption{The reaction term $f(u,v^*)$ for three values of $v^* \in [0.07,0.08,0.12]$. Note that $v^*=0.07$, indicated in red, is below the threshold for bistability. For all values of $v^*$ above the threshold value ($v^* \approx 0.073$), $f(u,v^*)$ has three roots.}
\end{figure}
\section{Constant wind}\label{constant wind}
In the case when the wind term is homogeneous in space, the traveling wave solutions described in \S \ref{traveling wave solutions} are heteroclinic connections between the fixed points found as zeros of \eqref{reaction term}. In the context of the ODE system \eqref{first order system} with such a wind term, $(u_0,0)$ and $(u_2,0)$ are saddles. These points define boundary conditions for the traveling wave in the $(u,s)$ system: as we take our traveling coordinate $z$ to negative infinity, the solution trajectory asymptotically approaches the point $(0,0)$. As we take our traveling coordinate $z$ to positive infinity, it approaches ($u_2$,0). \par
Classical theory tells us that solutions $\hat{u}$ to the corresponding second-order system \eqref{reduced TW equation} with a constant wind term exist only for a unique wave speed $c=\hat{c}$ (see \cite{evans2010partial}). We outline two approaches for finding these solutions. 
\subsection{Solutions in phase space}
The first method exploits the geometric structure described above. We reframe the first order system \eqref{first order system} as two initial value problems, one initiated near $(u_0,s_0)=(0,0)$ and one near $(u_0,s_0)=(u_2,0)$. The wave speed $c$ is a free parameter, and one can numerically integrate forwards from $(0,0)$ and backwards from $(u_2,0)$ to obtain solution trajectories for each initial value problem and choice of $c$. The $c$ value for which these solution trajectories coincide in a single, heteroclinic orbit is $\hat{c}$, the unique wave speed for which a profile solution exists. These solutions exist in phase space. However, reframing them in $(z,u(z))$ space---with the proper parameterization---allows us to reconstruct the profile solution. \\ 
\subsection{Solutions to the boundary value problem}
The second option bypasses the phase space point of view entirely and instead considers the second-order system \eqref{reduced TW equation} as a boundary value problem with boundary conditions given by 
\begin{align}\label{boundary conditions}
    \begin{split}
        &\lim_{ z \rightarrow -\infty}\hat{u}(z)=0 \\
        &\lim_{ z \rightarrow \infty}\hat{u}(z)=u_2
    \end{split}
\end{align}
The wave speed $c$ is a free parameter that is solved for alongside $u$. \par
The boundary value approach is efficient and allows us to easily specify the desired error tolerance when solving for $\hat{c}$. Either method allows us to find profile solutions and wave speeds that vary as a function of the initial fuel load parameter $v^*$.
A key point is that, for all values of $v^* \in [0,1]$, $\hat{c}<0$. This aligns with our physical understanding that the traveling front moves from regions of maximum temperature (representing burning fuel) to regions of ambient temperature (representing unburned fuel) as the fire progresses. The sign of $\hat{c}$ will also be an important point in our investigation of the system with a spatially dependent wind. 
\subsection{Stability}
We linearize the second order problem \eqref{traveling equation} about solutions to \eqref{reduced TW equation}, denoted $\hat{u}$. The resulting eigenvalue problem in linearized coordinates $p$ is 
\be \label{constant wind eval prob}
\lambda p =\mathcal{L}p:=p_{zz}+(c-w)p_z+f'(\hat{u})p
\ee
where 
\be \label{react prime}
f'(u(z))=\begin{cases}
\frac{v^*}{u^2} \exp\left(-\frac{1}{u}\right)-l \: &u(z)>0 \\
-l \: &u(z) = 0
\end{cases}
\ee
\par 
It is well known (see \cite{henry1981geometric}) that the boundary of the essential spectrum of the operator $\mathcal{L}$ associated with \eqref{constant wind eval prob} is given by the spectrum of the operator $\mathcal{L}^{\pm \infty}$, where the latter is determined by evaluating $\mathcal{L}$ at the profile solution $\hat{u}$ and taking the the limit as $z \rightarrow \pm \infty$. A straightforward algebraic argument demonstrates that the essential spectrum lies in the left half-plane. For details as to how this calculation works in a similar system, see \cite{ghazaryan2015stability}. There is also an eigenvalue at the origin reflective of the translational invariance of the traveling wave (see \cite{Sattinger1976on}): the monotonicity of the profile and an application of Sturm-Liouville theory tells us that the translational eigenvalue is the largest eigenvalue for any given solution (for details, see \cite{fife1977approach}). This in turn implies that the wave is spectrally stable. \par
The conclusion is that, in the spatially homogeneous wind case, profile solutions exist to \eqref{traveling equation} that move with a unique wave speed. These solutions are spectrally stable for all values of the wind speed and translates of the front are also solutions. We will see that, in the spatially dependent wind case, these results do not hold: solutions exist for a continuum of wave speeds and translational invariance is broken.
\begin{figure}[h]
    \centering
    \includegraphics[scale=1.2]{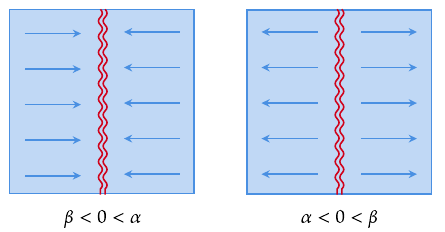}
    \caption{Two possibilities for the spatially dependent wind configuration. The first illustrates a convergent wind field and the second a divergent wind field. The fireline is indicated in red.}
\end{figure}
\section{Constructing the spatially dependent wind}
To introduce the switching phenomenon described in \S \ref{introduction}, we use a discontinuous step function whose magnitude is controlled by the parameters $\alpha$ and $\beta$:
\be\label{discontinuous wind}
w(z)=\begin{cases}
    \alpha \: \: &z <0 \\
    0 \: \: &z=0 \\
    \beta  \: \: &z>0 
\end{cases}
\ee 

This leaves us with two potential situations: either $\alpha>\beta$ or $\alpha<\beta$. As a special case of the first option, we might consider the situation in which $\beta<0<\alpha$, resulting in a wind field that converges at the fireline. Alternatively, as a special case of the second option, we consider $\alpha<0<\beta$, resulting in a divergent wind field. The first case describes the fire-induced wind introduced in \S \ref{introduction}. The second case is physically unlikely, but presents a relevant mathematical foil that will help us better understand the physical implications of our results.

\par
Recall that in the case of constant wind, traveling wave solutions exist for a unique $\hat{c}$. In the spatially dependent wind case, the situation is more complicated: solutions exist for a range of $c$ values, each corresponding to a different profile shape. We will use the structure outlined in \S \ref{constant wind} as a guide for our analysis of the spatially dependent wind case. 
\section{Solutions in phase space with a spatially dependent wind}\label{phase space solutions}
\begin{figure}[h]
    \centering
    \includegraphics[scale=.9]{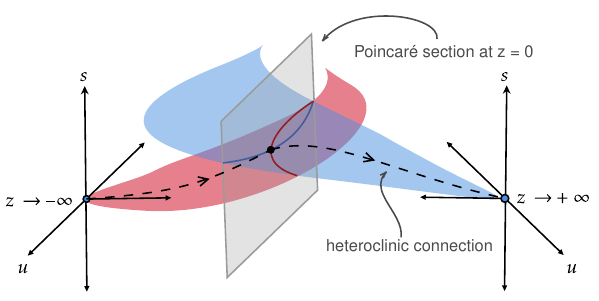}
    \caption{The 3d phase space described in \S 5. The blue plane is the stable manifold of the fixed point $(u_2,0)$ for the system at $+\infty$ and the  red plane is the unstable manifold of the fixed point $(0,0)$ for the system at $-\infty$. The solid blue and red curves indicate the intersections of these manifolds with the plane $z=0$. The point of intersection of the solid curves can be traced back across the unstable manifold and forward across the stable manifold to find the heteroclinic connection between $(0,0)$ and $(u_2, 0)$.}
    \label{3d phase space}
\end{figure}

With the nonconstant wind term \eqref{discontinuous wind} the ODE governing the front solutions \eqref{first order system} is now non-autonomous. As a first order autonomous system, this becomes 
\be\label{first order non-autonomous}
\begin{cases}
    &u'=s \\
    &s'=-(c-w(z))s-v^*r(u)+lu \\
    &z'=1
\end{cases}
\ee
The system \eqref{first order non-autonomous} has traveling wave solutions $\hat{u}(z) $ that exist for some bounded continuum of wave speeds $c$. 
There exists a set of orbits connecting the point $(u,s)=(0,0)$ as $z \rightarrow -\infty$ and the point $(u,s)=(u_2,0)$ as $z \rightarrow +\infty$.  We can think of the dynamics projected onto the plane $z'=1$ and therefore reduced to a 2d phase space in which both of these points are equilibria. The unstable manifold of the first fixed point is a 2d structure that intersects with the Poincaré section at $z=0$ in a 1d curve, as is the stable manifold of the second fixed point. The point of intersection of these 1d curves on the plane $z=0$ can then be traced back to the fixed points at $\pm \infty$ to form the heteroclinic connection, as illustrated in figure \ref{3d phase space}. \par
This view gives us two different systems for $z<0$ and $z>0$.\\
\\ For $z<0$:
\begin{align}\label{system 1}
\begin{split}
    u_-'&=s_- \\
    s_-'&=-c_1^*s_--f(u_-)
\end{split}    
\end{align}
where $c_1^*=c-\alpha$. For $z>0$:
\begin{align}\label{system 2}
\begin{split}
    u_+'&=s_+ \\
    s_+'&=-c_2^*s_+-f(u_+)
\end{split}    
\end{align}
where $c_2^*=c-\beta$.
Values of $c$ for which solution trajectories to the systems \eqref{system 1} and \eqref{system 2} intersect are values of $c$ for which a solution exists to the full system \eqref{first order system} with the wind switch. \par
To find these values we consider solution trajectories for \eqref{system 1} originating at the fixed point $(0,0)$ and tangent to the unstable subspace of that fixed point as well as trajectories for \eqref{system 2} originating at the fixed point $(u_2,0)$ tangent to the stable subspace  for that system. These trajectories are invariant manifolds for their respective system parameterized by the wave speed $c$, which is common to both systems. \par 
The point of intersection of the manifolds corresponds to the location of the wind switch and the point on the profile where the derivative $u'=s$ is maximized corresponds to the location of the fireline. We may pick our parameterization based on either of these points: a natural choice, since we define the wind switch as occurring at $z=0$ in \eqref{discontinuous wind}, is to fix the point of intersection of the manifolds at $z=0$.

\subsection{Finding the invariant manifolds and their intersections}
The fixed points of \eqref{system 1} and \eqref{system 2} are the same as the fixed points of \eqref{first order system} in the constant wind case and so are given by the zeros of \eqref{reaction term}. Again, they are saddles. Denoting a fixed point $(u,s)$ of either system as $(u^*,0)$, the matrix associated with the linearization of the system about the fixed point is 
\be 
M_{\pm}=
\begin{bmatrix}
    0 & 1 \\
    -f'(u^*) & -c^*
\end{bmatrix}
\ee
where $c^*=c^*_1$ or $c^*_2$ as appropriate and $f'$ is defined as in \eqref{react prime}.
For the fixed point $(u_2,0)$ at positive infinity, the stable eigenvalue of the linearization is 
\be
 \lambda_s=\frac{-c_2^*-\sqrt{{c^*_2}^2-4(f'(u_2))}}{2}
\ee
 and the corresponding eigenvector is 
 \be \label{stable eigenvector} \pm\vec{v}_s=\pm \begin{pmatrix}
     1 \\
     \lambda_s
 \end{pmatrix}
 \ee
 Similarly, for the fixed point $(0,0)$ at negative infinity, the unstable eigenvalue of the linearization is 
 \be\label{unstable eval from 0}
 \lambda_s=\frac{-c_1^*-\sqrt{{c^*_1}^2+4l}}{2}
 \ee
 and the corresponding eigenvector is 
 \be \label{unstable eigenvector} \pm\vec{v}_{us}=\pm\begin{pmatrix}
       1 \\
       \lambda_{us}
 \end{pmatrix}
 \ee
\\ We perturb the fixed point at $- \infty$ in the unstable direction to determine an initial condition. We then integrate forward from this initial condition to find the unstable manifold, treating our PDE as an initial value problem. We take a similar approach for the stable manifold originating from the fixed point at $+ \infty$. The range of $c$ values for which intersections exist depends on the relative values of the wind speed parameters $\alpha$ and $\beta$. These results are summarized in Proposition 1. 
\begin{prop}
If $\alpha > \beta$, the unstable manifold of the fixed point $(0,0)$ of \eqref{system 1} intersects the stable manifold of the fixed point $(u_2,0)$ of \eqref{system 2} when $c_1^*<\hat{c}$ and $c_2^*> \hat{c},$ where $\hat{c}$ is the unique wave speed for which solutions exist when $\alpha=\beta=0$ (or equivalently, the wind is constant across the spatial domain). If $\alpha < \beta,$ this intersection occurs for  $c_1^*>\hat{c}$ and $c_2^*< \hat{c}.$ This corresponds to $\hat{c}+\alpha < c < \hat{c}+\beta$ when $\alpha>\beta$ and $\hat{c}+\beta < c < \hat{c}+\alpha$ when $\alpha<\beta$. 
\end{prop}

\begin{proof}
We consider $\mathcal{W}^u_0(c)$, the unstable manifold originating at $(0,0)$ representing a solution trajectory of \eqref{system 1} with initial condition $(0,0)$ and speed $c$. $\mathcal{W}^u_0(c)$  is parameterized in forward ``time," meaning that the independent variable $z$ goes from $-\infty$ to $+\infty$ along the course of the trajectory. \par
 $\mathcal{W}^s_{u_2}(c)$ is the stable manifold originating from $(u_2,0)$ representing a solution trajectory of \eqref{system 2} with initial condition $(u_2,0)$ and speed $c$.  $\mathcal{W}^s_{u_2}(c)$  is parameterized in backward ``time," so the independent variable $z$ goes from $+\infty$ to $-\infty$ along the course of the trajectory. Note that when $c=\hat{c}$ these manifolds coincide, meaning $\mathcal{W}^s_{u_2}(\hat{c})=\mathcal{W}^u_{0}(\hat{c})$. \par
$\mathcal{W}^s_{u_2}(c)$ and $\mathcal{W}^u_{0}(c)$ are parameterized by the same wave speed $c$, which may or may not be a value of $c$ for which there is a solution for the entire system with the wind switch. For values of $c$ for which $\mathcal{W}^s_{u_2}(c)$ and $\mathcal{W}^u_{0}(c)$ intersect, there is such a solution. \par 
We also consider the energy of the system, defined by the Hamiltonian associated with \eqref{first order system} when $c^*=0$:
\begin{equation}\label{Hamiltonian}
    H(u,s)=\frac{1}{2}s^2+\int_0^u f(r) dr
\end{equation}
The derivative of \eqref{Hamiltonian} along trajectories is 
\begin{equation}\label{Hamiltonian deriv}
\frac{\mathrm{d} }{\mathrm{d} z}H(u(z),s(z))=-c^*s^2
\end{equation}
If $c^*=0$, $\frac{\mathrm{d} H }{\mathrm{d} z}$ is also zero along trajectories, meaning H is constant along trajectories, so level sets of \eqref{Hamiltonian} are solution trajectories of \eqref{first order system}. \par
The derivative of the Hamiltonian, \eqref{Hamiltonian deriv}, gives us information about the rate of change of energy along trajectories when $c^* \neq 0$. In particular, recall that  $\hat{c} < 0$ for $v^* \in [0,1]$. In the case of constant wind, this means that $c^*=\hat{c}<0 \implies \frac{\mathrm{d} H}{\mathrm{d} z} >0$, meaning energy is increasing along trajectories that are parameterized in forward $z$ and decreasing along trajectories parameterized in backward $z$. 

\begin{figure}[t]
\centering
\subfloat[]{\includegraphics[width=0.5\columnwidth]{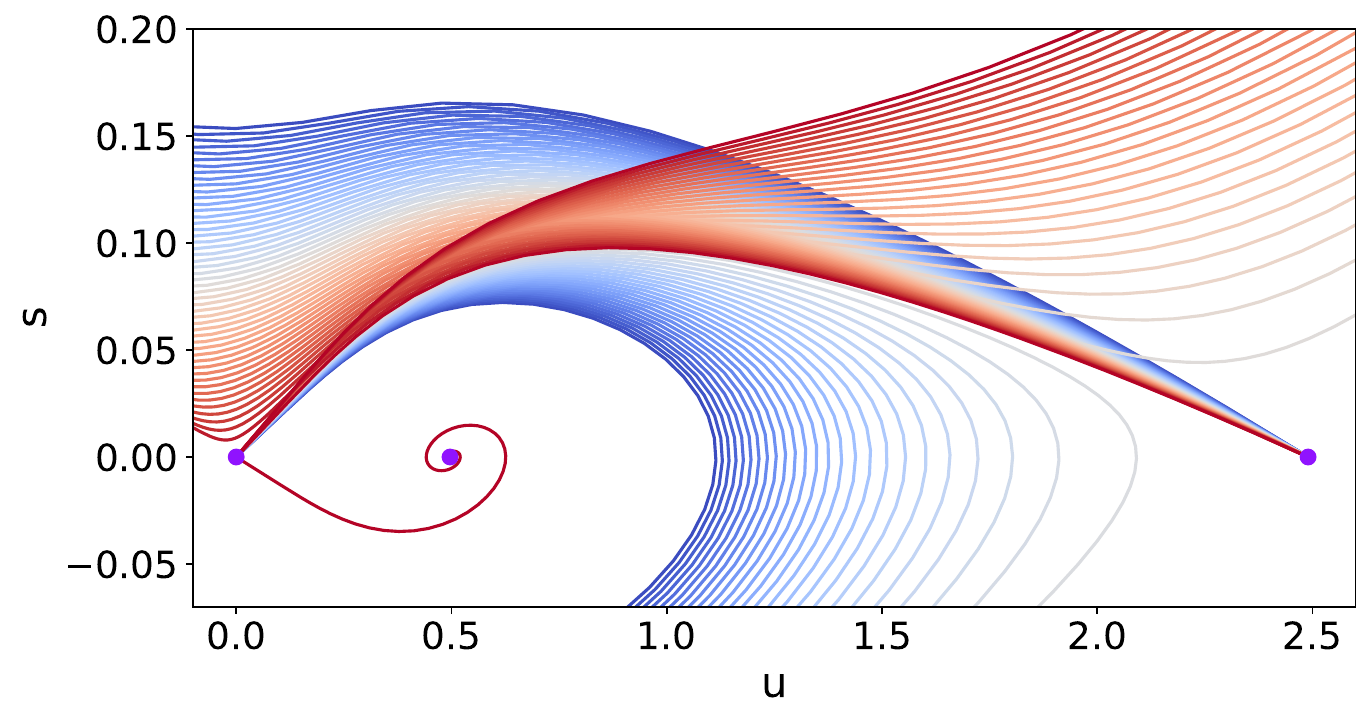}}
\subfloat[]{\includegraphics[width=0.5\columnwidth]{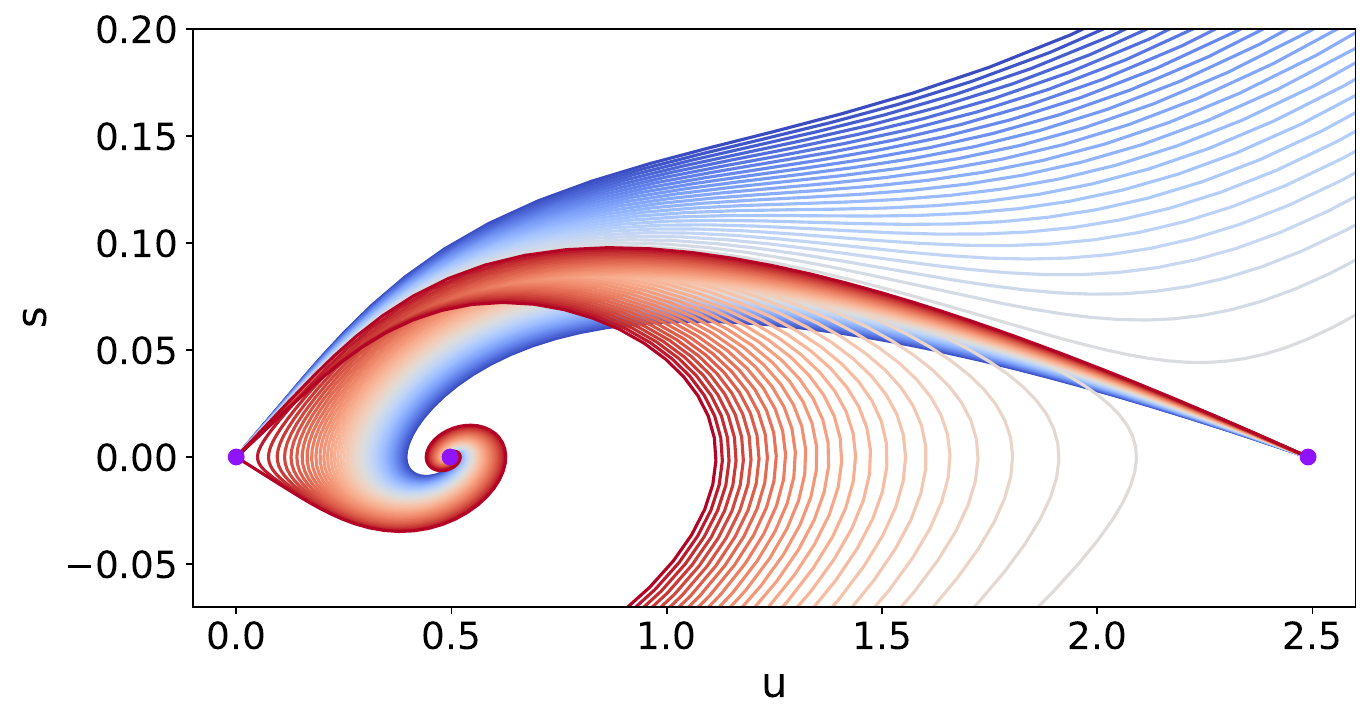}} \\

\caption{Invariant manifolds color-coded by wave speed value $c$. (a) depicts type 1 intersections for $c \in [-0.04547148,-0.12547148]$, while (b) depicts type 2 intersections for $c \in [-0.12547148,-0.04547148]$. The fixed points $(0,0)$, $(u_1,0)$ and $(u_2,0)$ are indicated in purple. Note the different qualitative nature of the intersections in (a) vs. those in (b).} 
\end{figure}
\subsection{Case 1: $\alpha > \beta$}
Consider $c_1^*=c-\alpha$. Assuming non-constant wind, either (a) $c^*_1>\hat{c}$ or (b) $c^*_1<\hat{c}$. If option (a) holds, $\frac{\mathrm{d} H}{\mathrm{d} z}(c_1^*)<\frac{\mathrm{d} H}{\mathrm{d} z} (\hat{c})$, meaning that energy increases less along the trajectories defined by $\mathcal{W}^u_0(c_1^*)$ than those defined by $\mathcal{W}^u_0(\hat{c})$. This, combined with the fact that the second component of the eigenvector from $(0,0)$ in the unstable direction is given by the unstable eigenvalue \eqref{unstable eval from 0} so that $c^*_1>\hat{c} \implies \lambda_{us}(c^*_1)<\lambda_{us}(\hat{c})$, means that trajectories $\mathcal{W}^u_0(c_1^*)$ will always lie below trajectories $\mathcal{W}^u_0(\hat{c})$ in phase space for case (a). If $\mathcal{W}^s_{u_2}(c_2^*)$ is to intersect $\mathcal{W}^u_0(c_1^*)$ it must also lie below $\mathcal{W}^u_0(\hat{c})=\mathcal{W}^s_{u_2}(\hat{c})$. Because $\mathcal{W}^s_{u_2}$ is parameterized in backwards $z$, this requires 
$\frac{\mathrm{d} H}{\mathrm{d} z}(c_2^*)>\frac{\mathrm{d} H}{\mathrm{d} z}(\hat{c})$ in forward $z$. This implies that $c_2*<\hat{c} \implies c-\beta<\hat{c} \implies c <\hat{c} + \beta$, and the condition on the stable manifold gives $c> \hat{c} +\alpha$. But this implies $\alpha < \beta$, contradicting our initial assumption. \par
So it must be that option (b) holds, meaning that $\frac{\mathrm{d}H}{\mathrm{d}z}(c_1^*)>\frac{\mathrm{d}H}{\mathrm{d}z}(\hat{c})$, meaning that energy increases more along the trajectories defined by $\mathcal{W}^u_0(c_1^*)$ than those defined by $\mathcal{W}^u_0(\hat{c})$. In this case, $ \lambda_{us}(c^*_1)>\lambda_{us}(\hat{c})$, which means that trajectories $\mathcal{W}^u_0(c_1^*)$ will always lie above trajectories $\mathcal{W}^u_0(\hat{c})$ in phase space. This implies that $c_2^*>\hat{c}$ and therefore $\hat{c}+\beta<c<\hat{c}+\alpha$, which is consistent with our initial assumption that $\alpha>\beta.$ \par
To demonstrate that $\mathcal{W}^u_0(\hat{c})$ and $\mathcal{W}^s_{u_2}(\hat{c})$ do in fact intersect for $c_1^*<\hat{c}<c_2^*$, consider the lines $u=0$ and $u=u_2$. We know that $\mathcal{W}^u$ originates at the point $(0,0)$ on the line $u=0$ and, because it is bounded, must intersect the line $u=u_2$. An analogous argument holds for $\mathcal{W}^s$ and the line $u=0$. \par
To see that the manifolds are bounded, we make the change of variables $s(z)=\exp(-c^*z)b(z)$. Then, the ODE for $s'$ in \eqref{first order system} becomes $b'=-f(u)$. Integrating, we find $$b(z)=-\int_{-\infty}^{\infty} f(u(z)) dz$$ The right hand side is bounded for $u \in [0, u_2]$ (see figure 1), so we may conclude that $s(z)$ is also bounded in that range. \par
Interpreting $\mathcal{W}^s$ and $\mathcal{W}^u$ as functions $\mathcal{W}:u \rightarrow s$, we further define $\mathbf{d}(u)=s_1(u)-s_2(u)$, where $s_2=\mathcal{W}^s(u)$ and $s_1=\mathcal{W}^u(u)$. The energy argument above tells us that $s_1$ and $s_2$ lie above the $u$ axis. So, we have:
\begin{align}\label{distance function}
\begin{split}
\mathbf{d}\left(u=0\right)&=0-s_2(0) < 0 \\
\mathbf{d}\left(u=u_2\right)&=s_1(u_2)-0 > 0 
\end{split}
\end{align}
Both $\mathcal{W}^u$ and $\mathcal{W}^s$ are continuous functions of $z$ because they are solution trajectories parameterized in $z$. Further, the fact that $u$ is a monotone bounded function of $z$ follows from that the fact that $s>0 \implies u' =s > 0$ and $s$ is bounded, as demonstrated in part (1). This, together with the fact that $u(z)$ satisfies the intermediate value property for $z \in (-\infty, \infty)$, gives us that $u$ is a continuous function of $z$. Therefore its inverse, $z(u)$, is also continuous. Because the composition of continuous functions is continuous, we conclude that both $\mathcal{W}^u$ and $\mathcal{W}^s$ are continuous functions of $u$. Finally, because $\mathbf{d}$ is a continuous function of $u$ and switches signs in the interval $[0, u_2]$, there must be a point $u^* \in [0, u_2]$ such that $\mathbf{d}\left(W^u_0(c_1^*),W^s_{u_2}(c_2^*),u^*\right)=0$, meaning that $\mathcal{W}^u_0$ and $\mathcal{W}^s_{u_2}$ intersect at this point.
\subsection{Case 2: $\alpha < \beta$}
By the same logic as case 1, we must have $c_2^*<\hat{c}$ and $c_1^*>\hat{c}$ in order for $\mathcal{W}^u_0$ and $\mathcal{W}^s_{u_2}$ to intersect, giving us $\hat{c}+\alpha<c<\hat{c}+\beta$. In this case, the sign of $\frac{\mathrm{d}H}{\mathrm{d}z}$ and the magnitude of the eigenvector in the stable direction tells us that $\mathcal{W}^u_0(c_1^*)$ and $\mathcal{W}^s_{u_2}(c_2^*)$ must both lie below $\mathcal{W}^s_{u_2}(\hat{c})=\mathcal{W}^u_{0}(\hat{c})$. \par
The fixed point $(u_1,0)$ is either an unstable node or an unstable spiral for the systems from both $-\infty$ and $+\infty.$ This implies that $(u,s)=(u_1,0)$ is the $\alpha$ limit set of the system from positive infinity, meaning that $\mathcal{W}^s_{u_2}$ must approach the point $(u_1,0)$ as $z \rightarrow -\infty$. \par
If $(u_1,0)$ is a node for this system, this  guarantees that the trajectory from $+\infty$ asymptotically approaches the point $(u_1,0)$. On the other hand, because $(u_1,0)$ is an unstable fixed point for the system from $-\infty$, which is parameterized in forward time, the $\omega$ limit set does not exist. Therefore, the trajectory from $-\infty$ diverges as $z \rightarrow +\infty$. \par
To see that solution trajectories for the systems at both $\pm\infty$ must cross the $u$ axis in specific regions, consider the vector field along that axis:
\begin{align*} 
u'&=0 \\
s'&= -f(u)
\end{align*}
where the graph of $-f(u)$ is positive for $0<u<u_1$ and negative for $u_1<u<u_2$. \par 
It follows that if $\mathcal{W}^u_0$ is to intersect the $u$ axis, it must do so in the region $u>u_1$ due to the direction of the vector field along the axis. The opposite is true for $\mathcal{W}^s_{u_2},$ as it is parameterized in backwards $z$: if it intersects the $u$ axis, it must do so in the region $u<u_1$. Because the solution trajectories to both systems are bounded and lie below the trajectory associated with $\hat{c}$, which intersects the $u$ axis, they must both intersect it as well: the above bounds on where they may do so means that they must also intersect each other. 
\end{proof}


\subsection{Stability properties of different intersection types}
We now have a natural means of classifying intersection types: either $\alpha < \beta,$ resulting in ``type 1" intersections, or $\beta<\alpha$, resulting in ``type 2" intersections. Numerical investigation (see figure 4) reveals that these two types of intersections are qualitatively different, and this qualitative difference translates to a difference in the stability properties of the intersections. These properties are summarized in Proposition 2.
\begin{prop} Intersections of type 1 correspond to stable fronts while intersections of type 2 correspond to unstable fronts. 
\end{prop}
\begin{proof}[Proof] 
\subsection{Case 1, $\alpha > \beta$: type 1 intersections}
Define the tangent vector associated with $\mathcal{W}_0^u$ at the point of intersection as $v_0^u$ and the tangent vector associated with $\mathcal{W}_{u_2}^s$ at the point of intersection as $v_0^s$. At the point of intersection, $v^u_0$ evolves according to the flow along $\mathcal{W}_{u_2}^s$. Call the resulting vector at the point $(u,s)=(u_2,0)$  $v^u_f$ and the tangent vector to the stable manifold at that point $v^s_f$. \par 
Note that, because $v^u_0$ and $v^s_0$ are tangent to the unstable and stable manifolds respectively, their components are determined by the vector field of the corresponding systems, namely \eqref{system 1} and \eqref{system 2}. The corresponding matrix is 
\be
A=\begin{pmatrix}
    u'_- & u'_+ \\
    s'_- & s'_+
\end{pmatrix}
\ee
and 
\be
\mathrm{det}(A)=u'_-s'_+-s'_-u'_+=(c_1^*-c_2^*)s^2
\ee
For intersections of type 1, $c_2^*<\hat{c}<c_1^*$ means $\mathrm{det}(A)>0$ and the tangent vectors therefore constitute a positively oriented basis for $\mathbb{R}^2$. 
Uniqueness of solutions to differential equations requires that the orientation between the vectors be preserved, so the orientation of the basis associated with $[v^u_f,v^s_f]$ must also be positive. \par
Next, consider the angular eigenvalue system, obtained by converting the eigenvalue problem associated with \eqref{traveling equation} to polar coordinates and noting that the equation for $\theta$ is independent from that for $r$, where $\omega$ is the streamwise eigenvalue:
\be\label{angular system}
\dot{\theta}=-(f'(\hat{u})-\omega)\cos^2\theta-(c-w(z))\cos\theta\sin\theta-\sin^2\theta
\ee
which has fixed points $\theta^{\pm}_{u/s}$ corresponding to the unstable/stable directions at $\pm \infty$, given by:
\begin{align}\label{theta fixed points}
\begin{split}
    \theta^+_{u/s} &=\arctan\left(\frac{-(c+\beta)\pm\sqrt{(c+\beta)^2+4(\omega-f'(u_2))}}{2}\right) \\
    \theta^-_{u/s} &=\arctan\left(\frac{-(c+\alpha)\pm\sqrt{(c+\alpha)^2+4(\omega+l)}}{2}\right) 
\end{split}
\end{align}
\label{sec:sample1}
If a given value of $\omega$ is an eigenvalue, the corresponding solution trajectory to \eqref{angular system} with initial condition $\theta_u^-$ must approach $\theta_f=\theta_s^++n\pi$ as $z \rightarrow \infty$ for some integer $n$.  \par
Fix $\omega=0$ and define $\theta_f$ as the angle between the positive $u$ axis and $v_f^s$. A standard argument shows that $0$ is not an eigenvalue, so we cannot have $\theta_f=\theta_s^+$. In fact, because of the positive orientation of $[v_f^u, v_f^s],$ it must be that $\theta_f>\theta_s^+.$ Because $\theta_f$ must be aligned with an invariant subspace at $+\infty$ it must be that $\theta_f=\theta_u^+$ when $\omega=0$. \par
Now, because $\theta_f$ is a piecewise monotone function of $\omega$, $w > 0 \implies \theta_f \geq \theta^+_u$, as varying $\omega$ induces counter-clockwise rotation.  But, because the vector field along the line $\theta=\frac{\pi}{2}$ is positive, meaning it points in the clockwise direction, $\theta$ is unable to rotate through this axis. We conclude that, for $\omega>0$, $\theta_f=\theta_u^+$ and, by definition, $\omega$ is not an eigenvalue. As there are no eigenvalues in the right half plane, type 1 intersections correspond to stable solutions. 
\begin{figure}[h]
    \centering
    \includegraphics[scale=1]{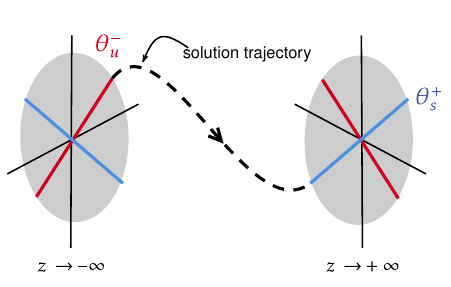}
    \caption{Phase portraits for the angular systems at $-\infty$ and $+\infty$ with a sample solution trajectory moving from $\theta_u^-$ to $\theta_s^+$. Unstable manifolds for each system are indicated in red and stable manifolds in blue.}
    \label{fig:enter-label}
\end{figure}
\subsection{Case 2, $\alpha < \beta$: type 2 intersections}
In this case, $c_1^*<\hat{c}<c_2^*$ means $\mathrm{det}(A)<0$ and the basis associated with the tangent vectors $[v_0^u,v_0^s]$ at the point of intersection is negatively oriented, as is the basis associated with $[v_f^u,v_f^s]$. \par
Again, fix $\omega=0.$ By the same logic as in case 1, $0$ is not an eigenvalue: the  orientation of $[v_f^u,v_f^s]$ means we must have $\theta_f=\theta_u^++\pi$. For $\omega >> 0$, $\theta_f$ approaches $\theta_u^+$ as the growing value of $\omega$ overcomes the effect of the nonlinear reaction term. In doing so, it must pass through $\theta_s^+ + \pi$ for some value of $\omega>0$. Therefore, for intersections of type 2, there is at least one value of $\omega$ in the right half plane. The corresponding solution is unstable.
\end{proof}\par
The geometric approach described in this section allows us to reconstruct profile solutions and identify the intersection types corresponding to stable profiles. However, we are left with a range of stable---and therefore physically viable---fronts and no means of picking an unique front solution from the continuum of possibilities. In the next two sections, we develop tools to identify the preferred solution.  

\section{Profile solutions to the boundary value problem with a spatially dependent wind}\label{spatialy dependent bv problem}
In \S \ref{selection mechanism}, we will seek to determine the preferred wave speed and corresponding solution by finding the minimizer of the largest eigenvalue associated with each wave speed/solution pair. This approach requires that we have access to very accurate numerical representations of the solutions whose existence we demonstrated in Proposition 1. As in the constant wind case, we accomplish this by framing our problem as a boundary value problem in which the wave speed $c$ is a free parameter. \par 
The system \eqref{first order system} is defined for $z \in \left(-\infty, \infty\right)$, so we look for profile solutions by solving the two-point boundary value problem on this domain, where the problem is given by \eqref{first order system} with wind term \eqref{discontinuous wind} and boundary conditions \eqref{boundary conditions}. We also have a condition on the derivative of the profile, which takes the form of a pulse:
\be\label{8}
\lim_{z \rightarrow \pm\infty}s =0 \\
\ee
We follow the technique outlined in \cite{barker2018evans}, slightly modified to allow for a discontinuous wind. We employ projective boundary conditions
\begin{equation}
    P_{\pm}\left(U(\pm L)-U_{\pm}\right)=0
\end{equation}
where $U=\left(u,s\right)$, $U_+=(u_2,0)$, $U_-=(0,0)$ and we take $\pm L$ to be sufficiently large so as to approximate numerical infinity. \par 
There exists a one-parameter family of solutions indexed by the wave speed $c$. To pick a single solution from this family, we specify a phase condition
\be \label{phase condition}
\text{phase}(\gamma)=\gamma u_2+(1-\gamma) u_0
\ee
for $\gamma \in [0,1]$. This phase condition serves as an additional boundary condition specifying the spatial location of the center of the profile, which corresponds exactly to the position of the fireline as determined by the point where the derivative $u'=s$ is at a maximum. Imposing a phase condition allows us to leave the wave speed $c$ as an unknown to be solved for, whereas in \S 7 we specified $c$ and found the intersections of the resulting invariant manifolds in $\left(u,s(u)\right)$ space.  \par
A number of pre-packaged solvers exist for two-point BVPs, so it is convenient to transform our three-point BVP in to a two-point BVP by doubling the size of the system and halving the domain. Lastly, to account for the unknown wave speed parameter, we augment the system with the additional relation $c'=0$.
Then, \eqref{first order system} becomes 
\begin{align}\begin{split}
    \label{6} 
  u_0' &=s_0 \\
  s_0' &=-(c-w(z))u_0-f(u_0)  \\
  u_1' &= -s_1 \\
  s_1' &=(c-w(z))u_1+f(u_1) \\
  c' &=0
\end{split}
\end{align}
for $z \in [0,L].$
To close the system, we introduce a set of matching conditions at the $z=0$ boundary, resulting in five total boundary conditions for our five-dimensional system. The resulting solutions are smooth profiles asympotically connecting $u=0$ to $u=u_2$, as expected.

\begin{figure}[h]
\centering
\includegraphics[scale=0.5]{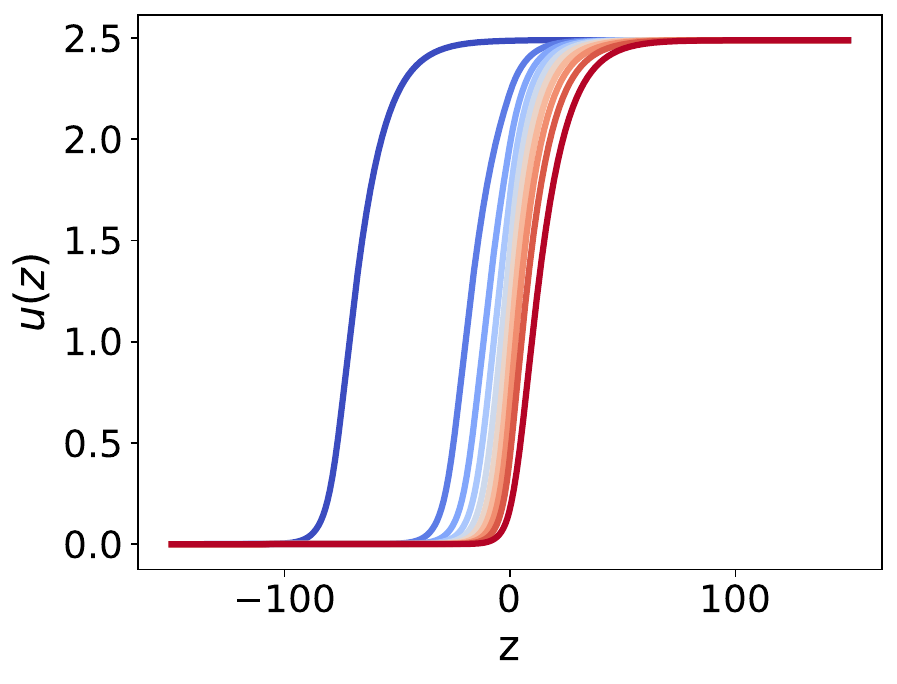} 
\caption{Profile solutions found using the method described in \S \ref{spatialy dependent bv problem}, for $v^*=0.1$, $l=0.02688$, $\alpha=0.05$, $\beta=-\alpha$, color-coded by wave speed $c \in [-0.03547, -0.13546]$. In this case, $u_2 \approx 2.5$.}
\label{figure: profile solutions}
\end{figure}

\begin{figure}[h]
\centering
\includegraphics[scale=0.5]{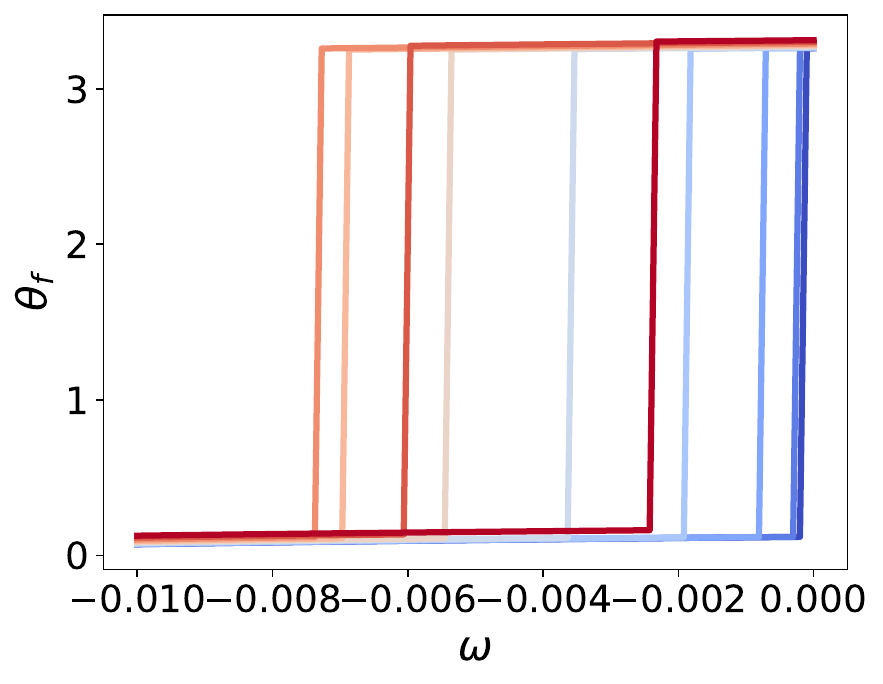}
\caption{The accumulated angle as a function of $\omega$ color-coded by wave speed $c \in [-0.03547, -0.13546]$. Vertical jumps indicate the $\omega$ value that is an eigenvalue for that value of $c$ and corresponding solution.} 
\end{figure}

\section{Stability as a selection mechanism for the preferred wave}\label{selection mechanism}

The results presented in Propositions 1 and 2 give us a means of understanding the existence and stability of traveling front solutions to \eqref{nondimensional equation} with a spatially dependent wind. In particular, these results affirm our physical intuition: for a divergent wind field---corresponding to $\alpha < \beta$ and type 2 intersections---the front is unstable, whereas for a convergent wind---representative of the fire-induced wind described in \S\ref{introduction} and corresponding to $\alpha>\beta$ and type 1 intersections---the front is reinforced by the wind field and the solution is stable. \par
We expect for stable fronts---those whose speeds fall in the range $\hat{c}+\alpha < c < \hat{c} + \beta$--- to persist in nature, so we focus the remainder of our analysis on traveling fronts resulting from type 1 intersections. The solutions we found in \S \ref{spatialy dependent bv problem} for a wind term satisfying this inequality exist for a range of phase conditions, just as intersections of the invariant manifolds discussed in \S \ref{phase space solutions} exist for a range of wave speeds. However, without a means of selecting a single wave speed from the continuum of possibilities, the solution for the system with a spatially dependent wind term is not unique. \par 
A reasonable means of identifying the preferred solution is to identify the solution whose largest (meaning least negative) streamwise eigenvalue $\omega$ is minimized across possible phase conditions or, equivalently, wave speeds. Because perturbations to this solution will decay the fastest, it is in a sense the ``most stable" of the possible solutions. Out of the continuum of possible stable solutions, this spectrally preferred solution corresponds to the traveling front that will persist in nature. \par 
To find the largest eigenvalue for each profile solution found in \S \ref{spatialy dependent bv problem}, we again turn to the angular formulation given in \eqref{angular system}. Solutions take the form of trajectories that originate on the unstable manifold of the fixed point $(0,0)$ for the system at $-\infty$ and terminate on the stable manifold of the fixed point $(u_2,0)$ for the system at $+\infty$, where the two systems are differentiated by the wind term, as in \eqref{system 1} and \eqref{system 2}. \par
If some value of the eigenvalue parameter $\omega$ is in fact an eigenvalue, the corresponding trajectory in the angular system $\theta(z)$ will satisfy the following conditions: 
\be \label{eval condition}
\begin{split}
\lim_{z \rightarrow -\infty}\theta& =\theta_u^- \\ 
\lim_{z \rightarrow +\infty}\theta&=\theta^+_s
\end{split}
\ee
for $\theta_u^-$ and $\theta_s^+$  defined as in \eqref{theta fixed points}.
\par

To determine if a particular value of $\omega$ satisfies this condition, we employ a shooting method, solving \eqref{angular system} by integrating forward in $z$ with initial condition $\theta_0=\theta_u^-$ and $\hat{u}$ given by the solutions to the BVP discussed in \S \ref{spatialy dependent bv problem}. \par
For each candidate value of $\omega$, we consider $\theta_f(\omega)$, always initializing at $\theta_0=\theta_u^-.$ The final value of the $\theta(z)$ can be thought of as the angle ``accumulated" over the course of the trajectory. A jump of size $\pi$ in the accumulated angle indicates that the corresponding $\omega$ value is an eigenvalue, as such a jump can only be achieved if $\theta_f$ has moved from $\theta_u^+$ to $\theta_u^++\pi$, passing through $\theta_s^+$ or $\theta_s^++\pi$ in the process and therefore satisfying condition \eqref{eval condition}. \par 
\begin{figure}
\centering
\subfloat[]{\includegraphics[width=0.5\columnwidth]{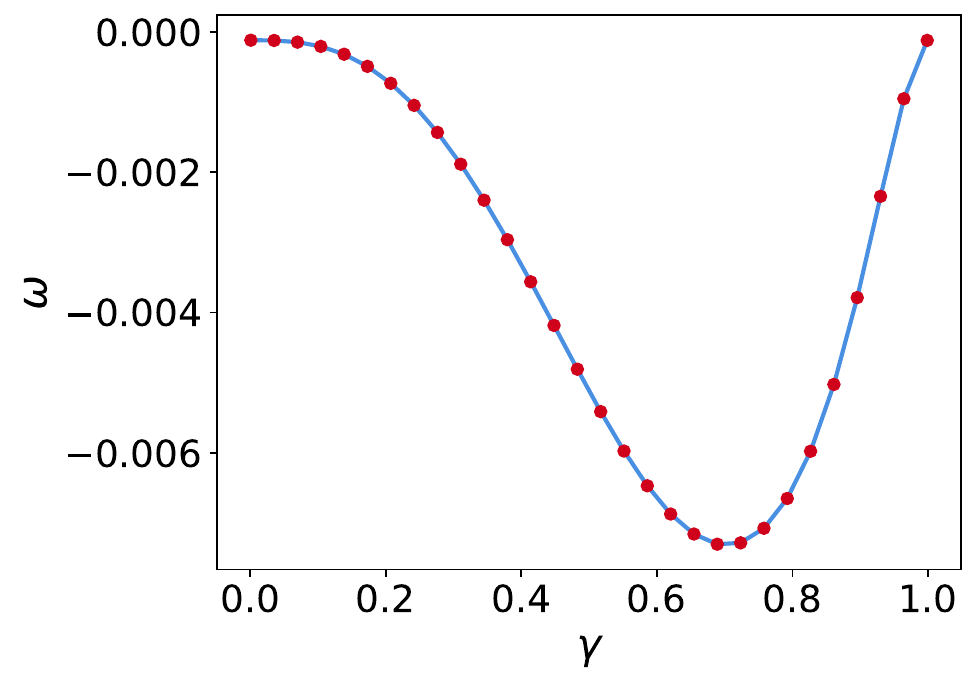}} 
\subfloat[]{\includegraphics[width=0.5\columnwidth]{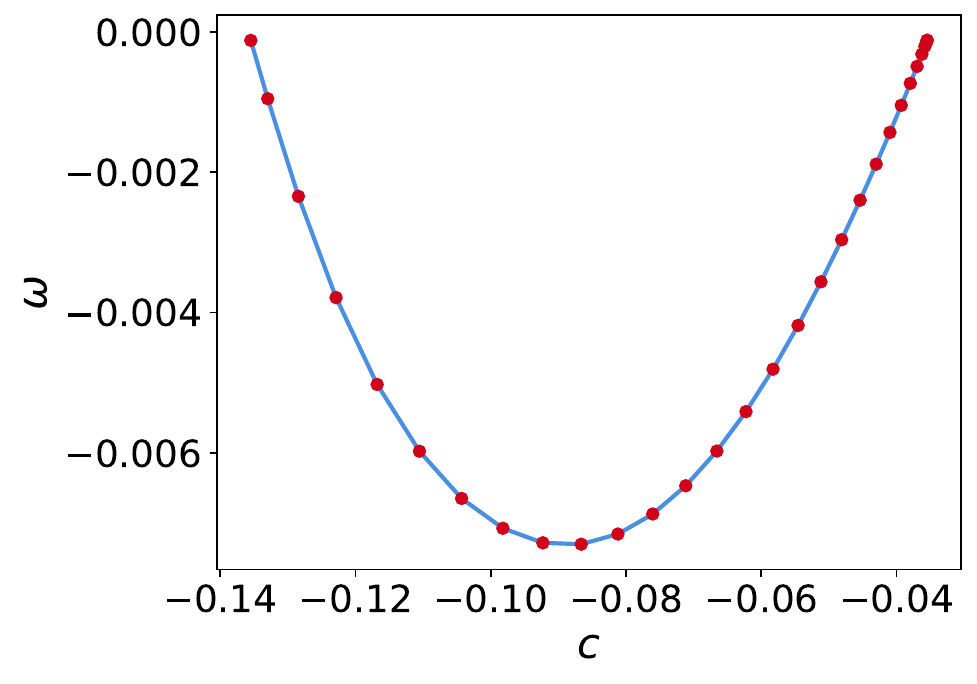}} 
\caption{The largest eigenvalue $\omega$ as a function of the phase $\gamma$ in figure (a) and the wave speed $c$ in figure (b). Both views demonstrate a clear minimum and an inverted parabolic structure.} \label{mincurve}
\end{figure}
The jump we find for each solution for the parameter ranges indicated in figure \ref{figure: profile solutions} is never larger than $\Delta\theta=\pi$. An application of Sturm-Liouville theory tells us that, for each solution, this eigenvalue is the first in a string of eigenvalues ordered $\omega_0>\omega_1> \dots > - \infty$ and, therefore, the largest. \par
Finally, we use a root finding algorithm to identify the exact values of $\omega$ at which these jumps occur. Examining the largest eigenvalue as a function of the wave speed (alternatively, the phase) for solutions associated with that wave speed (phase) gives us a curve with a clear minimum: see figure \ref{mincurve}. The solution corresponding to this minimum eigenvalue is the most stable, and therefore preferred, solution.
\par Note that both eigenvalue curves are essentially inverted parabolas, with $\omega$ only approaching $0$ as $\gamma$ and $c$ approach the limits of their respective ranges. This occurs for the limiting cases of $c=\hat{c}+\beta$ and $c=\hat{c}+\alpha$, either end of the range of allowable speeds found in \S \ref{phase space solutions}. It is only for these limiting cases that an eigenvalue exists at or very near zero: for all other allowable wave speeds, there is no zero eigenvalue and therefore no translational invariance. \par
\begin{figure}[h!]
\centering
\subfloat[]{\includegraphics[scale=0.4]{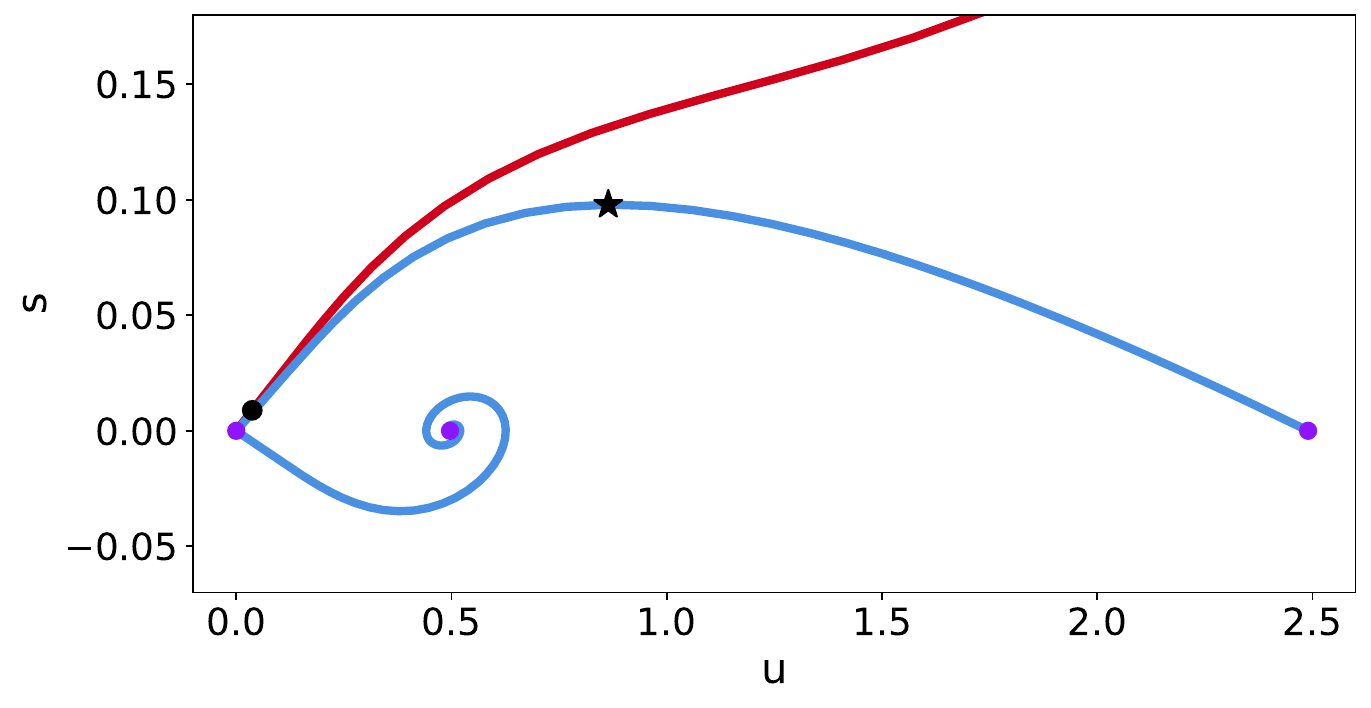}} \\
\subfloat[]{\includegraphics[scale=0.4]{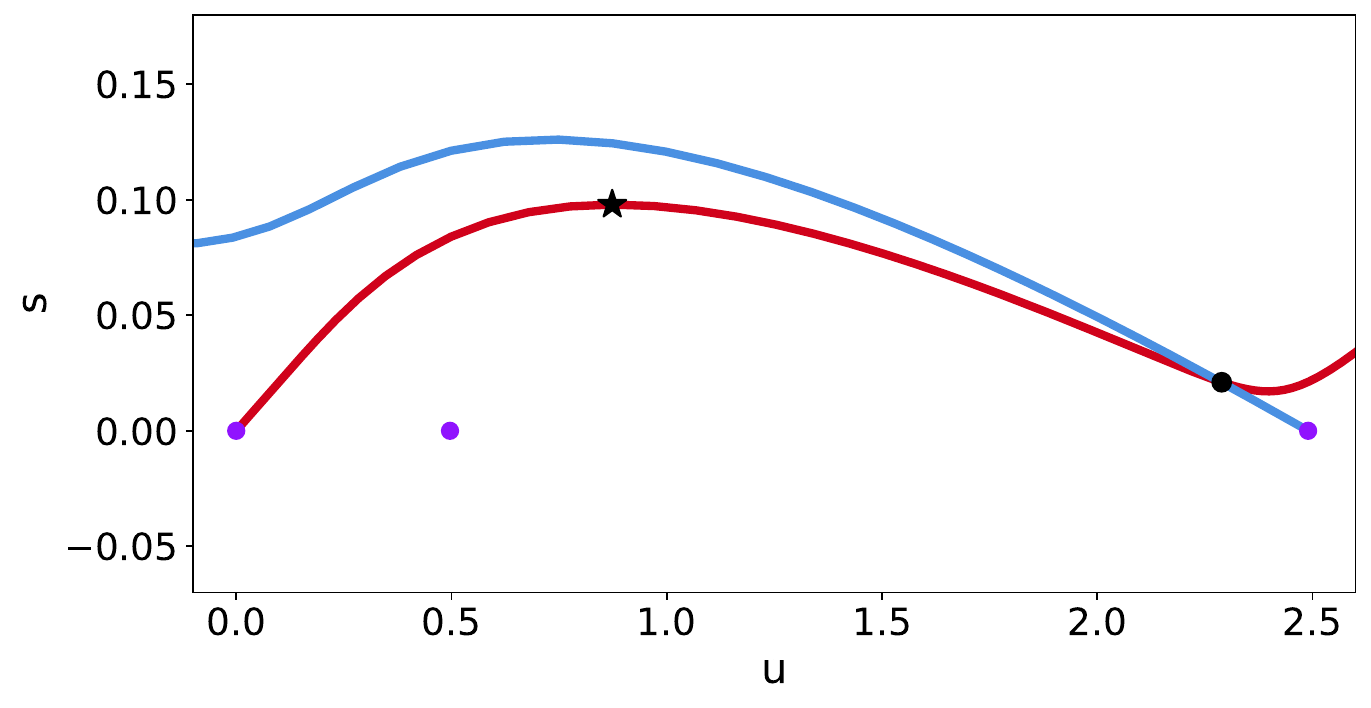}}\\ 
\subfloat[]{\includegraphics[scale=0.4]{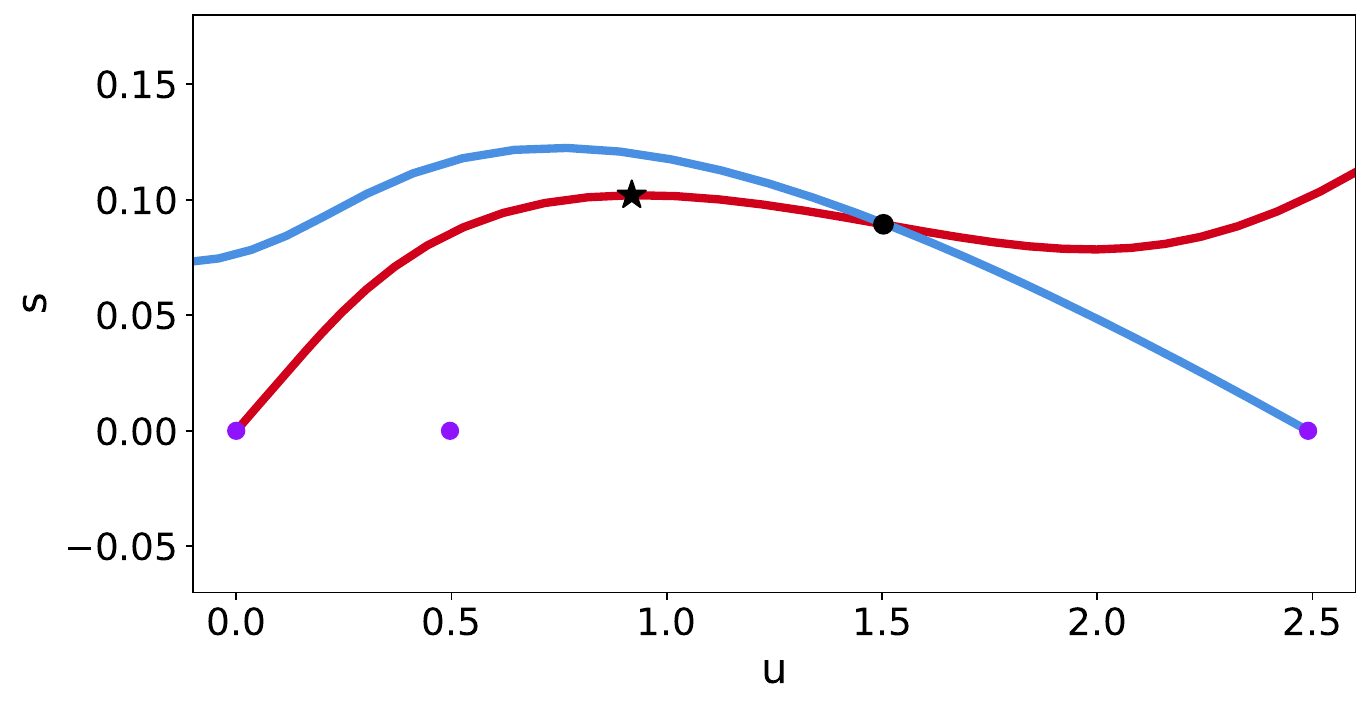}}\\
\caption{Intersecting invariant manifolds for two wave speeds on either end of the allowed range as well as the preferred wave speed. The fireline occurs at the maximum value of $s$ along the portion of the trajectory corresponding to the solution and is indicated by the black star. The wind switch occurs at the point of intersection of the manifolds and is indicated by the black dot. All intersections are type 1. (a) shows the intersection when $c$ is very close to $\hat{c}+\beta$. The signed distance between the fireline and the wind switch is $\mathbf{d}=-0.82683$. (b) shows the intersection when $c$ is very close to $\hat{c}+\alpha$, with $\mathbf{d}=1.41559$. (c) shows the intersection for the preferred wave speed, $c=-0.08995$, with $\mathbf{d}=0.58459.$}
\label{cases}
\end{figure}

 For the phase space trajectories corresponding to both limiting situations and the preferred case, as well as the distance between the fireline and wind switch in each case, see figure \ref{cases}. For wind speeds near the ends of the allowable range, the wind switch occurs very close to $z=\pm \infty$ and far away from the fireline. The preferred solution corresponds to a wave speed of $c=-0.08995$. \par 
 Notably, for the preferred solution, the wind switch is closer to the fireline than either of the limiting cases as well as being spatially ``ahead." Specifically, if we compute the signed distance between the $u$ value of the wind switch and that of the fireline as $\mathbf{d}=u_{windswitch}-u_{fireline}$, we find $\mathbf{d}= 0.58459$ for the preferred wave speed (as illustrated in image (c) in figure \ref{cases}). We parameterize the front solutions so that $z=0$ corresponds to the fireline. Therefore $\mathbf{d}>0$ means that, for the preferred solution, the wind switch occurs ahead of the fireline in a manner consistent with the physics of air entrainment.

\section{Conclusion} 
Understanding the effects of changes in local environmental variables such as wind on the spatial structure of a propagating fireline is key to better understanding fire behavior more generally. Mathematically, the existence and stability of traveling wave solutions in a system with spatial discontinuities, such as those imposed by the wind switch studied here, is an intriguing extension to classical theory. In this paper, we explored the effects of an external forcing term describing the fire-induced wind on the existence of profiles in the temperature representing fire fronts and their resilience to spatial perturbations.  \par
Traveling wave solutions are heteroclinic connections between fixed points of the corresponding system---in this case, in $\mathbb{R}^2.$ As such, problems concerning their existence and stability lend themselves well to geometric dynamical systems approaches. We extended the classical existence argument involving the coincidence of invariant manifolds to a spatially discontinuous system, in which existence must be demonstrated through the intersection of invariant manifolds because of the difference between the systems at $\pm \infty$. We found that intersections, and therefore solutions, exist only for a bounded continuum of wave speeds. Moreover, the interval of wave speeds for which solutions exist is dependent on the relative magnitude of the wind speed before and after the point of discontinuity. The relationship between the wind speed parameters also determines which of two qualitatively different categories the intersection falls in to: type 1 or type 2. This classification is tied to the stability of the fronts and allows us to filter out those wind configurations corresponding to unstable---and therefore physically unviable---fronts.\par
As a means of identifying the spectrally preferred front and its speed, we calculated the magnitude of the largest eigenvalue across the range of wave speeds for which stable fronts exist. This led us to two physical conclusions: firstly, the largest eigenvalue is smaller than zero, as $\omega$ only approaches zero very near to the endpoint of the allowed range of wave speeds and the preferred speed is near the center of the range. The lack of a zero eigenvalue means that translational invariance is broken, so the fronts are localized in space. Secondly, the distance between the fireline and the wind switch is relatively close for the preferred solution, with the wind switch occurring ahead of the fireline. \par
Ultimately, we are able to use the mathematical criteria for existence and stability of front solutions to determine constraints on the physical configuration of the wind field and its spatial relationship to the fireline. We modeled only the temperature of the fire layer and imposed the combined ambient and fire-induced wind as an external forcing. We found that, for the spectrally preferred solution, the wind switch occurs ahead of the fireline. Despite not explicitly modeling the wind field, this key physical takeaway is consistent with the physics of air entrainment seen in fires at all scales.
\\
\\
\textbf{CRediT authorship contribution statement}\\
\\
\textbf{Olivia Chandrasekhar}: Writing -- original draft, Writing -- review \& editing, Formal analysis, Conceptualization, Visualization, Funding acquisition. \textbf{Christopher Jones}: Writing -- review \& editing, Formal analysis, Conceptualization, Supervision. \textbf{Blake Barker}: Writing -- review \& editing, Software, Validation. \textbf{Rod Linn}: Writing -- review \& editing, Supervision, Funding acquisition.\\
\\
\textbf{Declaration of competing interests} \\
\\
Blake Barker serves on the early career editorial board of Physica D: Nonlinear Phenomena. Chris Jones serves on the honorary editorial board of Physica D: Nonlinear Phenomena. All other authors declare that they have no known competing financial interests or personal relationships that could have appeared to influence the work reported in this paper.
\\
\\
\textbf{Acknowledgements} \\
O.C. was supported by the National Science Foundation Graduate Research Fellowship under Grant No. DGE-2439854. O.C. and R.L. were supported by the LDRD Grant ``Fighting Fire with Fire: Enabling a Proactive Approach to Wildland Fire" No. 20220024DR. C.J. was supported by Office of Naval Research Grant N00014-24-1-2198. \\
\\
\textbf{Supporting materials} \\
\noindent The code for profile generation, as described in \S \ref{spatialy dependent bv problem}, uses \textsc{STABLAB} \cite{barker2015stablab}. The code for the analysis and figures in this paper is available as part of the \textsc{STABLAB} repository at \url{https://github.com/nonlinear-waves/stablab_matlab/tree/master/streamwise_wildfire} and in a stand-alone format at \url{https://github.com/o-chandra/streamwise_wildfire}.

 \bibliographystyle{elsarticle-num} 
 \bibliography{main}

\begin{thebibliography}{10}
\expandafter\ifx\csname url\endcsname\relax
  \def\url#1{\texttt{#1}}\fi
\expandafter\ifx\csname urlprefix\endcsname\relax\def\urlprefix{URL }\fi
\expandafter\ifx\csname href\endcsname\relax
  \def\href#1#2{#2} \def\path#1{#1}\fi

\bibitem{mandel2008wildland}
J.~Mandel, L.~S. Bennethum, J.~D. Beezley, J.~L. Coen, C.~C. Douglas, M.~Kim, A.~Vodacek, A wildland fire model with data assimilation, Mathematics and Computers in Simulation 79~(3) (2008) 584--606.
\newblock \href {https://doi.org/https://doi.org/10.1016/j.matcom.2008.03.015} {\path{doi:https://doi.org/10.1016/j.matcom.2008.03.015}}.

\bibitem{norbury1988travelling1}
J.~Norbury, A.~M. Stuart, Travelling combustion waves in a porous medium. part i-existence, SIAM Journal on Applied Mathematics 48~(1) (1988) 155--169.
\newblock \href {https://doi.org/https://doi.org/10.1137/0148007} {\path{doi:https://doi.org/10.1137/0148007}}.

\bibitem{norbury1988travelling2}
J.~Norbury, A.~M. Stuart, Travelling combustion waves in a porous medium. part ii—stability, SIAM Journal on Applied Mathematics 48~(2) (1988) 374--392.
\newblock \href {https://doi.org/https://doi.org/10.1137/0148019} {\path{doi:https://doi.org/10.1137/0148019}}.

\bibitem{ghazaryan2015stability}
A.~Ghazaryan, S.~Lafortune, P.~Mclarnan, Stability analysis for combustion fronts traveling in hydraulically resistant porous media, SIAM Journal on Applied Mathematics 75~(3) (2015) 1225--1244.
\newblock \href {https://doi.org/https://doi.org/10.1137/140981204} {\path{doi:https://doi.org/10.1137/140981204}}.

\bibitem{hilton2018incorporating}
J.~Hilton, A.~Sullivan, W.~Swedosh, J.~Sharples, C.~Thomas, Incorporating convective feedback in wildfire simulations using pyrogenic potential, Environmental Modelling \& Software 107 (2018) 12--24.
\newblock \href {https://doi.org/https://doi.org/10.1016/j.envsoft.2018.05.009} {\path{doi:https://doi.org/10.1016/j.envsoft.2018.05.009}}.

\bibitem{lautenberger2013wildland}
C.~Lautenberger, Wildland fire modeling with an {Eulerian} level set method and automated calibration, Fire Safety Journal 62 (2013) 289--298.
\newblock \href {https://doi.org/https://doi.org/10.1016/j.firesaf.2013.08.014} {\path{doi:https://doi.org/10.1016/j.firesaf.2013.08.014}}.

\bibitem{mallet2009modeling}
V.~Mallet, D.~E. Keyes, F.~E. Fendell, Modeling wildland fire propagation with level set methods, Computers \& Mathematics with Applications 57~(7) (2009) 1089--1101.
\newblock \href {https://doi.org/https://doi.org/10.1016/j.camwa.2008.10.089} {\path{doi:https://doi.org/10.1016/j.camwa.2008.10.089}}.

\bibitem{atchley2021effects}
A.~L. Atchley, R.~Linn, A.~Jonko, C.~Hoffman, J.~D. Hyman, F.~Pimont, C.~Sieg, R.~S. Middleton, Effects of fuel spatial distribution on wildland fire behaviour, International Journal of Wildland Fire. 30: 179-189. 30 (2021) 179--189.
\newblock \href {https://doi.org/https://doi.org/10.1071/WF20096} {\path{doi:https://doi.org/10.1071/WF20096}}.

\bibitem{Linn2005numerical}
R.~R. Linn, P.~Cunningham, Numerical simulations of grass fires using a coupled atmosphere--fire model: Basic fire behavior and dependence on wind speed, Journal of Geophysical Research: Atmospheres 110~(D13) (2005).
\newblock \href {https://doi.org/https://doi.org/10.1029/2004JD005597} {\path{doi:https://doi.org/10.1029/2004JD005597}}.

\bibitem{canfield2014numerical}
J.~M. Canfield, R.~R. Linn, J.~A. Sauer, M.~Finney, J.~Forthofer, A numerical investigation of the interplay between fireline length, geometry, and rate of spread, Agricultural and Forest Meteorology 189-190 (2014) 48--59.
\newblock \href {https://doi.org/https://doi.org/10.1016/j.agrformet.2014.01.007} {\path{doi:https://doi.org/10.1016/j.agrformet.2014.01.007}}.

\bibitem{nelson2012entrainment}
R.~M. Nelson, B.~W. Butler, D.~R. Weise, Entrainment regimes and flame characteristics of wildland fires, International Journal of Wildland Fire 21~(2) (2012) 127.
\newblock \href {https://doi.org/https://doi.org/0.1071/WF10034} {\path{doi:https://doi.org/0.1071/WF10034}}.

\bibitem{linn2012using}
R.~Linn, J.~Canfield, P.~Cunningham, C.~Edminster, J.-L. Dupuy, F.~Pimont, Using periodic line fires to gain a new perspective on multi-dimensional aspects of forward fire spread, Agricultural and Forest Meteorology 157 (2012) 60--76.
\newblock \href {https://doi.org/https://doi.org/10.1016/j.agrformet.2012.01.014} {\path{doi:https://doi.org/10.1016/j.agrformet.2012.01.014}}.

\bibitem{dipierro2024simple}
S.~Dipierro, E.~Valdinoci, G.~Wheeler, V.-M. Wheeler, A simple but effective bushfire model: Analysis and real-time simulations, SIAM Journal on Applied Mathematics 84~(4) (2024) 1504--1514.
\newblock \href {https://doi.org/https://doi.org/10.1137/24M1644596} {\path{doi:https://doi.org/10.1137/24M1644596}}.

\bibitem{sandstede2002stability}
B.~Sandstede, Stability of travelling waves, in: Handbook of dynamical systems, Vol.~2, Elsevier, 2002, pp. 983--1055.

\bibitem{evans2010partial}
L.~C. Evans, Partial differential equations, American Mathematical Society, Providence, R.I., 2010.

\bibitem{henry1981geometric}
D.~Henry, Geometric Theory of Semilinear Parabolic Equations, Springer, 1981.

\bibitem{Sattinger1976on}
D.~Sattinger, On the stability of waves of nonlinear parabolic systems, Advances in Mathematics (1976).
\newblock \href {https://doi.org/https://doi.org/10.1016/0001-8708(76)90098-0} {\path{doi:https://doi.org/10.1016/0001-8708(76)90098-0}}.

\bibitem{fife1977approach}
J.~M. Paul C.~Fife, The approach of solutions of nonlinear diffusion equations to travelling front solutions, Archive for Rational Mechanics and Analysis (1977).
\newblock \href {https://doi.org/https://doi.org/10.1007/BF00250432} {\path{doi:https://doi.org/10.1007/BF00250432}}.

\bibitem{barker2018evans}
L.~J. E. e.~a. Barker~B., Humpherys~J., Evans function computation for the stability of travelling waves, Philos Trans A Math Phys Eng Sci. (2018).
\newblock \href {https://doi.org/https://doi.org/10.1098/rspa.1997.0062} {\path{doi:https://doi.org/10.1098/rspa.1997.0062}}.

\bibitem{barker2015stablab}
B.~Barker, J.~Humpherys, J.~Lytle, K.~Zumbrun, \href{https://github.com/nonlinear-waves}{Stablab: A matlab-based numerical library for evans function computation} (2015).
\newline\urlprefix\url{https://github.com/nonlinear-waves}

\end{thebibliography}

\end{document}